\documentclass[fluids,article,submit,oneauthor,10pt,a4paper]{mdpi}

\preto{\abstractkeywords}{\nolinenumbers}

\firstpage{1}
\articlenumber{x}
\doinum{10.3390/------}
\pubvolume{xx}
\pubyear{2019}
\copyrightyear{2019}
\externaleditor{Academic Editor: name}
\history{Received: date; Accepted: date; Published: date}

\Title{Turbulence Intensity Scaling: A Fugue}

\Author{Nils T. Basse $^{1}$\orcidA{}}

\AuthorNames{Nils T. Basse}

\address{%
$^{1}$ \quad Elsas v\"{a}g 23, 423 38 Torslanda, Sweden; nils.basse@npb.dk}

\abstract{We study streamwise turbulence intensity definitions using smooth- and rough-wall pipe flow measurements made in the Princeton Superpipe. Scaling of turbulence intensity with the bulk (and friction) Reynolds number is provided for the definitions. The turbulence intensity scales with the friction factor for both smooth- and rough-wall pipe flow. Turbulence intensity definitions providing the best description of the measurements are identified. A procedure to calculate the turbulence intensity based on the bulk Reynolds number (and the sand-grain roughness for rough-wall pipe flow) is outlined.}


\begin{document}
%%%%%%%%%%%%%%%%%%%%%%%%%%%%%%%%%%%%%%%%%%
%% Only for the journal Gels: Please place the Experimental Section after the Conclusions

%%%%%%%%%%%%%%%%%%%%%%%%%%%%%%%%%%%%%%%%%%
\section{Introduction}

The turbulence intensity (TI) is of great importance in e.g. industrial fluid mechanics where it can be used for computational fluid dynamics (CFD) simulations as a boundary condition \cite{ansys_a}. The TI is at the center of the fruitful junction between fundamental and industrial fluid mechanics.

This paper contains an extension of the TI scaling research in \cite{russo_a} (smooth-wall pipe flow) and \cite{basse_a} (smooth- and rough-wall pipe flow). As in those papers, we treat streamwise velocity measurements from the Princeton Superpipe \cite{hultmark_a,smits_a}. The measurements were done at low speed in compressed air with a pipe radius of about 65 mm. Details on e.g. the bulk Reynolds number range and uncertainty estimates can be found in \cite{hultmark_a}. Other published measurements including additional velocity components can be found in \cite{willert_a} (smooth-wall pipe flow) and \cite{schultz_a,flack_a} (smooth- and rough-wall pipe flow).

Our approach to streamwise TI scaling is global averaging; physical mechanisms include separate inner- and outer-region phenomena and interactions between those \cite{marusic_a}. Here, the inner (outer) region is close to the pipe wall (axis), respectively.

The local TI definition (see e.g. Figure 9 in \cite{laufer_a}) is:

\begin{equation}
I(r)=\frac{v_{\rm RMS}(r)}{v(r)},
\end{equation}

\noindent where $r$ is the radius ($r=0$ is the pipe axis and $r=R$ is the pipe wall), $v(r)$ is the local mean streamwise flow velocity and $v_{\rm RMS}(r)$ is the local root-mean-square (RMS) of the turbulent streamwise velocity fluctuations. Using the radial coordinate $r$ means that outer scaling is employed for the position. As was done in \cite{russo_a,basse_a}, we use $v$ as the streamwise velocity (in much of the literature, $u$ is used for the streamwise velocity).

The measurements in \cite{laufer_a} were on turbulent flow in a two-dimensional channel and similar work for pipe flow was published in \cite{laufer_b}.

In this paper, we study TI defined using a global (radial) averaging of the streamwise velocity fluctuations. The mean flow is either included in the global averaging or as a reference velocity. This covers the majority of the standard TI definitions.

There is a plethora of TI definitions, which is why we use the term fugue in the title. This is inspired by \cite{touponce_a}, where Frank Herbert's {\it Dune} novels \cite{herbert_a} are interpreted as "an ecological fugue".

The ultimate purpose of our work is to be able to present a robust and well-researched formulation of the TI; an equivalent TI in the presence of shear flow. Our work is not adding significant knowledge of the fundamental processes \cite{marusic_b,alfredsson_a,monkewitz_a}, but we need to understand them in order to use them as a foundation for the scaling expressions.

The main contributions of this paper compared to \cite{basse_a} are:
\begin{itemize}
  \item The introduction of additional definitions of the TI
  \item Log-law fits in addition to power-law fits
  \item New findings on the rough pipe friction factor behaviour of the Princeton Superpipe measurements
\end{itemize}

Furthermore, we include a discussion on the link between the TI and the friction factor \cite{orlandi_a,basse_a} in the light of the Fukagata-Iwamoto-Kasagi (FIK) identity \cite{fukagata_a}.

Our paper is organized as follows: In Section \ref{sec:vel_defs}, we introduce the velocity definitions. These are used in Section \ref{sec:ti_defs} to define various TI expressions. In Section \ref{sec:ti_scal} we present scaling laws using the presented definitions. We discuss our findings in Section \ref{sec:disc} and conclude in Section \ref{sec:conc}.

\section{Velocity Definitions}
\label{sec:vel_defs}

The friction velocity is:

\begin{equation}
v_{\tau}=\sqrt{\tau_w/\rho},
\end{equation}

\noindent where $\tau_w$ is the wall shear stress and $\rho$ is the fluid density.

The area-averaged (AA) velocity of the turbulent fluctuations is:

\begin{equation}
\langle v_{\rm RMS} \rangle_{\rm AA} = \frac{2}{R^2} \times \int_0^R v_{\rm RMS}(r)r{\rm d}r
\end{equation}

\begin{figure}
%\vspace{0.5cm}
\centering
\includegraphics[width=10cm]{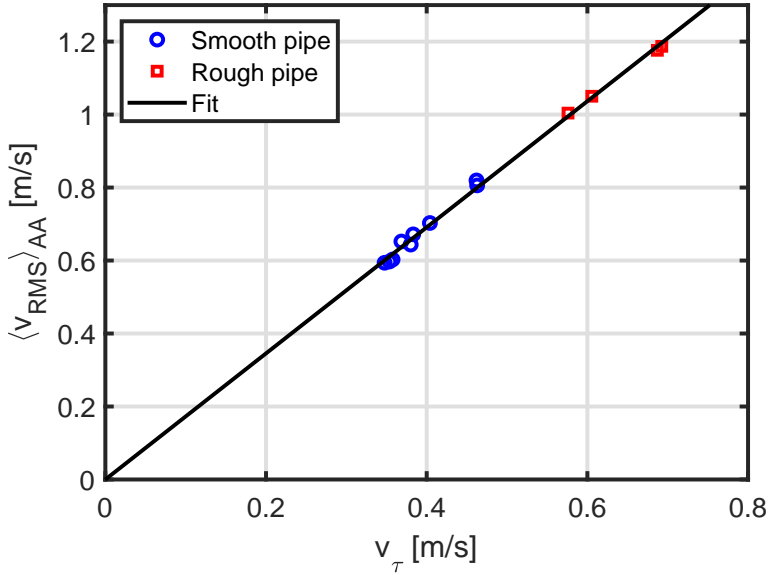}
\caption{Relationship between $v_{\tau}$ and $\langle v_{\rm RMS} \rangle_{\rm AA}$.}
\label{fig:AA_rms_fric_vels}
\end{figure}

The fit between $v_{\tau}$ and $\langle v_{\rm RMS} \rangle_{\rm AA}$ is shown in Figure \ref{fig:AA_rms_fric_vels}:

\begin{equation}
\langle v_{\rm RMS} \rangle_{\rm AA} = 1.7277 \times v_{\tau}
\label{eq:fric_rms}
\end{equation}

The velocity on the pipe axis is the centerline (CL) velocity:

\begin{equation}
v_{\rm CL} = v(r=0)
\end{equation}

The (area-averaged) mean velocity is given by:

\begin{equation}
v_m = \frac{2}{R^2} \times \int_0^R v(r)r{\rm d}r
\end{equation}

\begin{figure}
%\vspace{0.5cm}
\centering
\includegraphics[width=10cm]{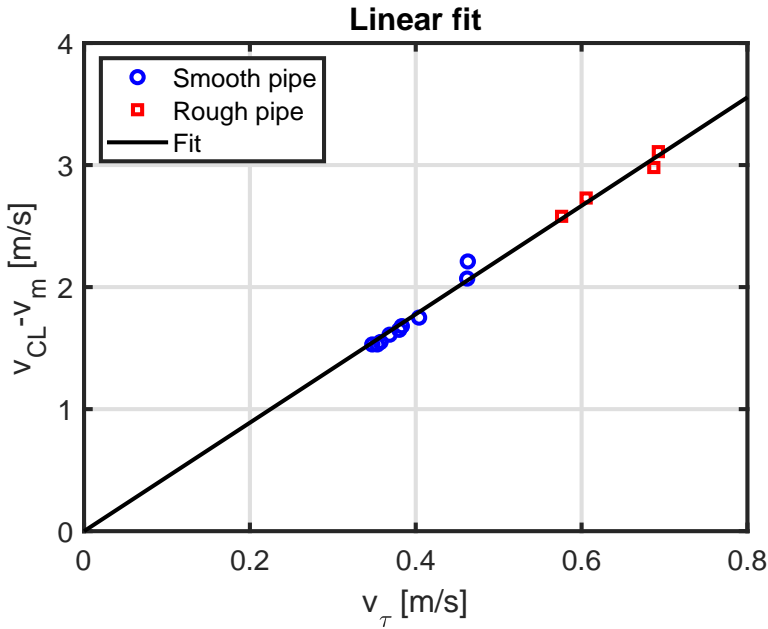}
\caption{Relationship between $v_{\tau}$ and $v_{\rm CL}-v_m$.}
\label{fig:cl_v_m}
\end{figure}

The difference between the centerline and the mean velocity scales with the friction velocity. The corresponding fit is shown in Figure \ref{fig:cl_v_m}:

\begin{equation}
\label{eq:fit_cl_v_m}
v_{\rm CL}-v_m=4.4441 \times v_{\tau},
\end{equation}

\noindent where the fit constant is close to the value of 4.28 \cite{gersten_a} found using earlier Princeton Superpipe measurements \cite{smits_b}.

%\vspace{0.5cm}

\section{Turbulence Intensity Definitions}
\label{sec:ti_defs}

\subsection{Local Velocity Definitions}

The arithmetic mean (AM) definition is:

\begin{equation}
I_{\rm Pipe~area,~AM} = \frac{1}{R} \int_0^R \frac{v_{\rm RMS}(r)}{v(r)} {\rm d}r
\end{equation}

The area-averaged definition is:

\begin{equation}
I_{\rm Pipe~area,~AA} = \frac{2}{R^2} \int_0^R \frac{v_{\rm RMS}(r)}{v(r)}r {\rm d}r
\end{equation}

In \cite{basse_a} (Equation (9)), $\langle v_{\rm RMS} \rangle$ was defined as the product of $v_m$ and $I_{\rm Pipe~area,~AA}$. Comparing the resulting Equations (11) and (12) in \cite{basse_a} to the current Equation (\ref{eq:fric_rms}), we find a difference of less than 5\% ($9/5$ compared to 1.7277).

Finally, the volume-averaged (VA) definition (inspired by the FIK identity) is:

\begin{equation}
I_{\rm Pipe~area,~VA} = \frac{3}{R^3} \int_0^R \frac{v_{\rm RMS}(r)}{v(r)}r^2 {\rm d}r
\end{equation}

\subsection{Reference Velocity Definitions}

As mentioned in the Introduction, we use outer scaling for the radial position $r$. For the TI, we separate the treatment to inner and outer scaling below, see e.g. \cite{alfredsson_a}.

\subsubsection{Inner Scaling}

For inner scaling, we define the TI using $v_{\tau}$ as the reference velocity:

\begin{equation}
I_{\tau} = \frac{\langle v_{\rm RMS} \rangle_{\rm AA}}{v_{\tau}} = 1.7277,
\end{equation}

\noindent where the final equation is found using Equation (\ref{eq:fric_rms}).

The square of the local version of this, $I_{\tau}^2(r) = \frac{v_{\rm RMS}(r)^2}{v_{\tau}^2}$,  is often used as the TI in the literature, see also Section 5.1 in \cite{basse_a}. This is the normal streamwise Reynolds stress normalised by the friction velocity squared.

\begin{figure}
%\vspace{0.5cm}
\centering
\includegraphics[width=10cm]{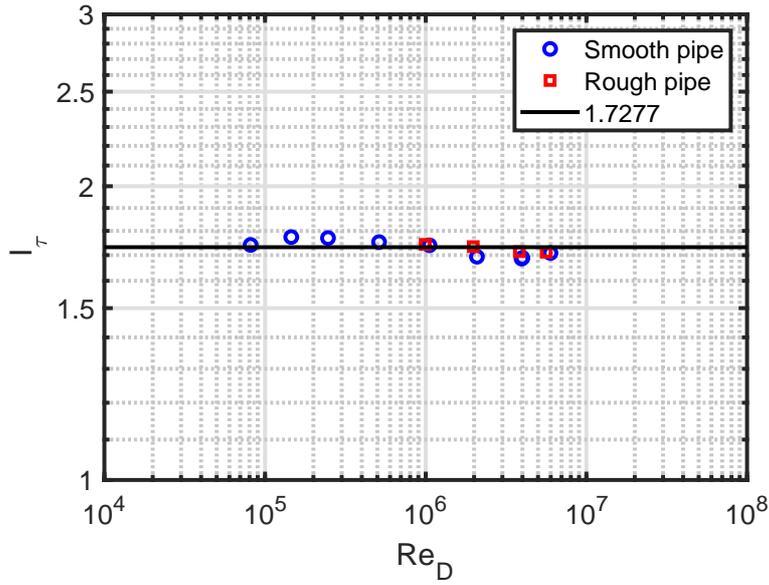}
\caption{$I_{\tau}$ as a function of $Re_{\rm D}$.}
\label{fig:AA_I_tau_smooth_rough}
\end{figure}

$I_{\tau}$ is shown in Figure \ref{fig:AA_I_tau_smooth_rough}; no scaling is observed with $Re_{\rm D}$, the bulk Reynolds number:

\begin{equation}
Re_{\rm D}=\frac{Dv_m}{\nu_{\rm kin}},
\end{equation}

\noindent where $D=2R$ is the pipe diameter and $\nu_{\rm kin}$ is the kinematic viscosity.

\subsubsection{Outer Scaling}

For outer scaling, we use either $v_m$ or $v_{\rm CL}$ as the reference velocity.

The TI using $v_m$ as the reference velocity is:

\begin{equation}
I_m = \frac{\langle v_{\rm RMS} \rangle_{\rm AA}}{v_m}
\label{eq:I_m}
\end{equation}

And finally, the TI using $v_{\rm CL}$ as the reference velocity \cite{alfredsson_a} is:

\begin{equation}
I_{\rm CL} = \frac{\langle v_{\rm RMS} \rangle_{\rm AA}}{v_{\rm CL}}
\end{equation}

\section{Turbulence Intensity Scaling Laws}
\label{sec:ti_scal}

The Princeton Superpipe measurements and the TI definitions in Section \ref{sec:ti_defs} are used to create the TI data points. Thereafter we fit the points using the power-law fit:

\begin{equation}
\label{eq:pow-fit}
Q_{\rm Power-law~fit} (x)=a \times x^b,
\end{equation}

\noindent and the log-law fit:

\begin{equation}
\label{eq:log-fit}
Q_{\rm Log-law~fit} (x)=c \times \ln (x) + d
\end{equation}

Here, $a$, $b$, $c$ and $d$ are constants. $Q$ is the quantity to fit and $x$ is a corresponding variable. We first apply the two fits using $Q=I$ and $x=Re_{\rm D}$.

The log-law fit is obtained by taking the (natural) logarithm of the power-law fit. The reason we use these two fits is that they have been discussed in the literature \cite{barenblatt_a,zagarola_a} as likely scaling candidates. We apply the two fits to the measurements and calculate the resulting root-mean-square deviations (RMSD) between the fits and the measurements. A small RMSD means that the fit is closer to the measurements.

Note that the smooth pipe $Re_{\rm D}$ measurement range is much larger than the rough pipe $Re_{\rm D}$ measurement range: A factor of 74 (9 points) compared to a factor of 6 (4 points). The consequence is a major uncertainty in the rough pipe results, e.g. (i) fits and (ii) extrapolation.

It is also important to be aware that we only have two sets of measurements with the following sand-grain roughnesses $k_s$:
\begin{itemize}
  \item Smooth pipe: $k_s = 0.45~\mu$m \cite{zagarola_b}
  \item Rough pipe: $k_s = 8~\mu$m \cite{langelandsvik_a} (see the related discussion in Section \ref{subsec:ff})
\end{itemize}

\begin{figure}
\centering
\includegraphics[width=7cm]{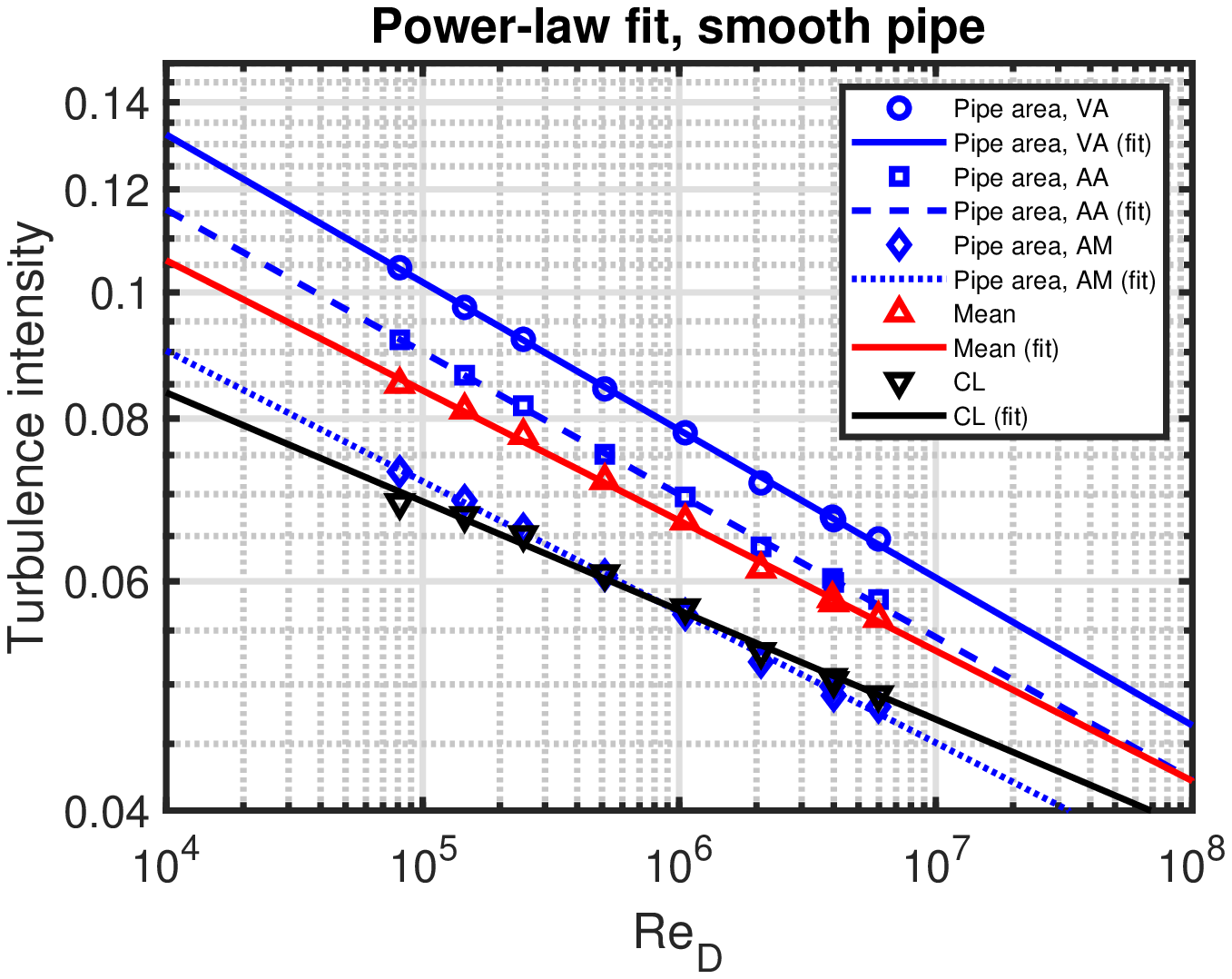}
\includegraphics[width=7cm]{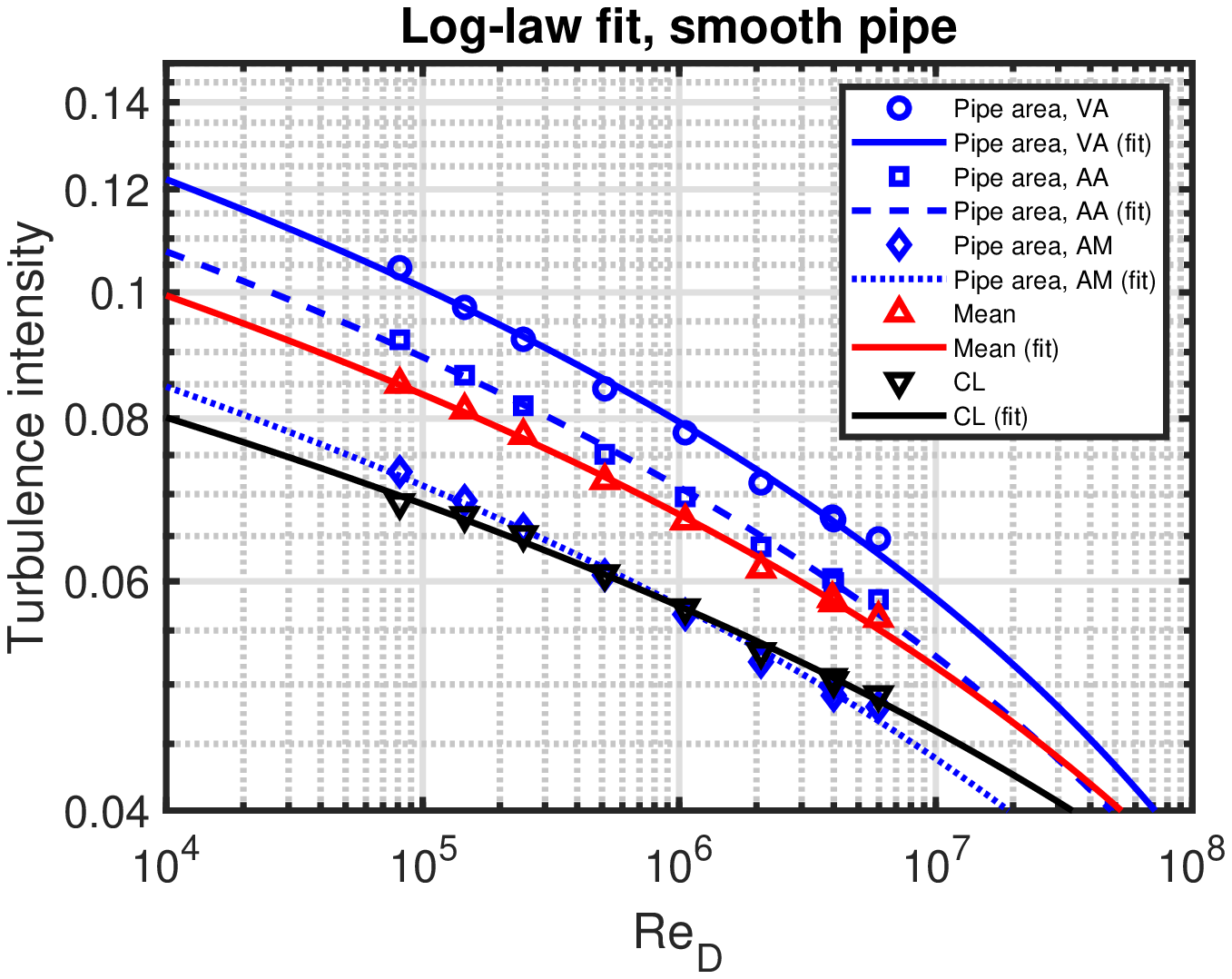}
\caption{Smooth pipe turbulence intensity as a function of $Re_{\rm D}$, left: Power-law fit, right: Log-law fit.}
\label{fig:I_smooth_rough}
\end{figure}

\begin{figure}
\centering
\includegraphics[width=7cm]{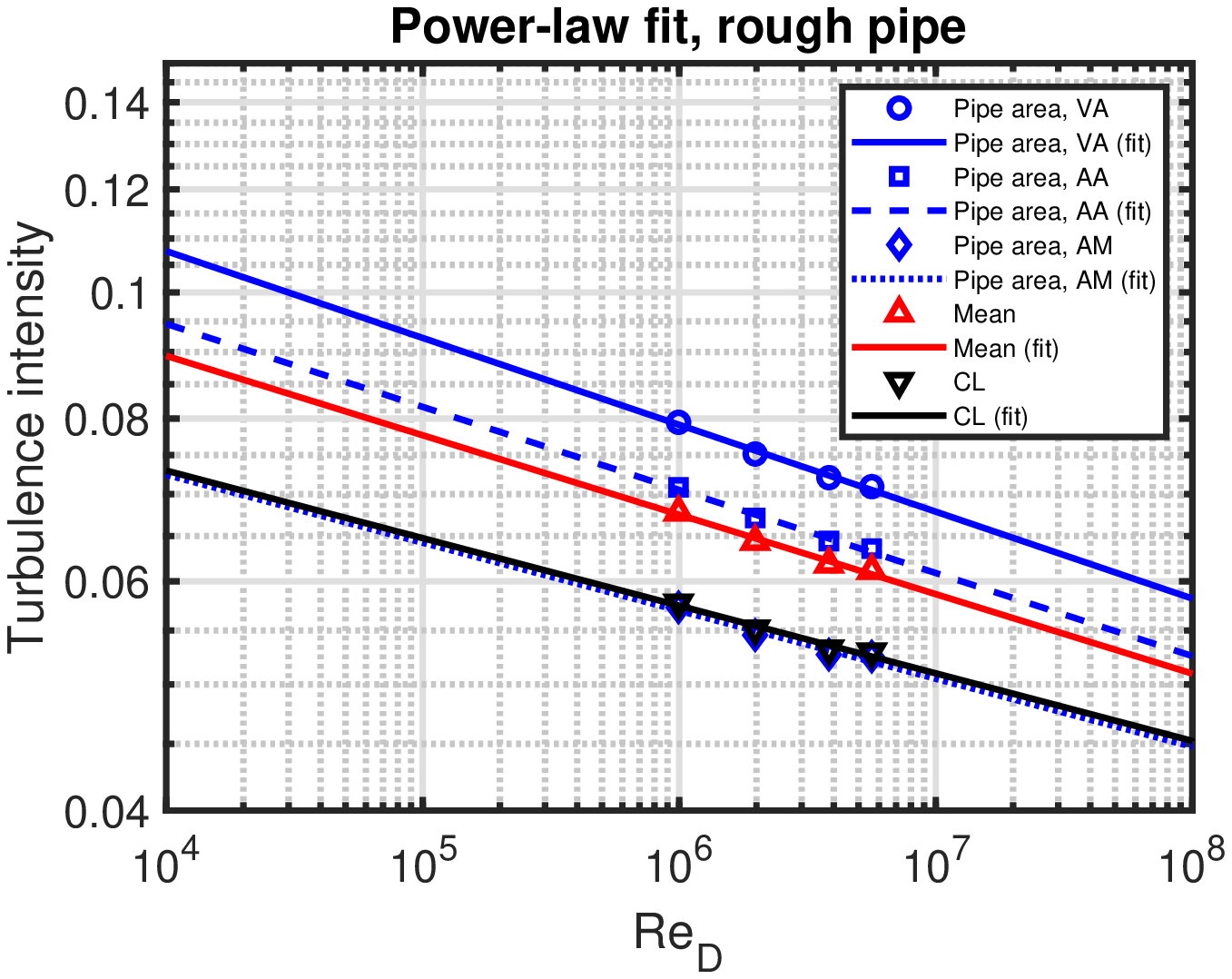}
\includegraphics[width=7cm]{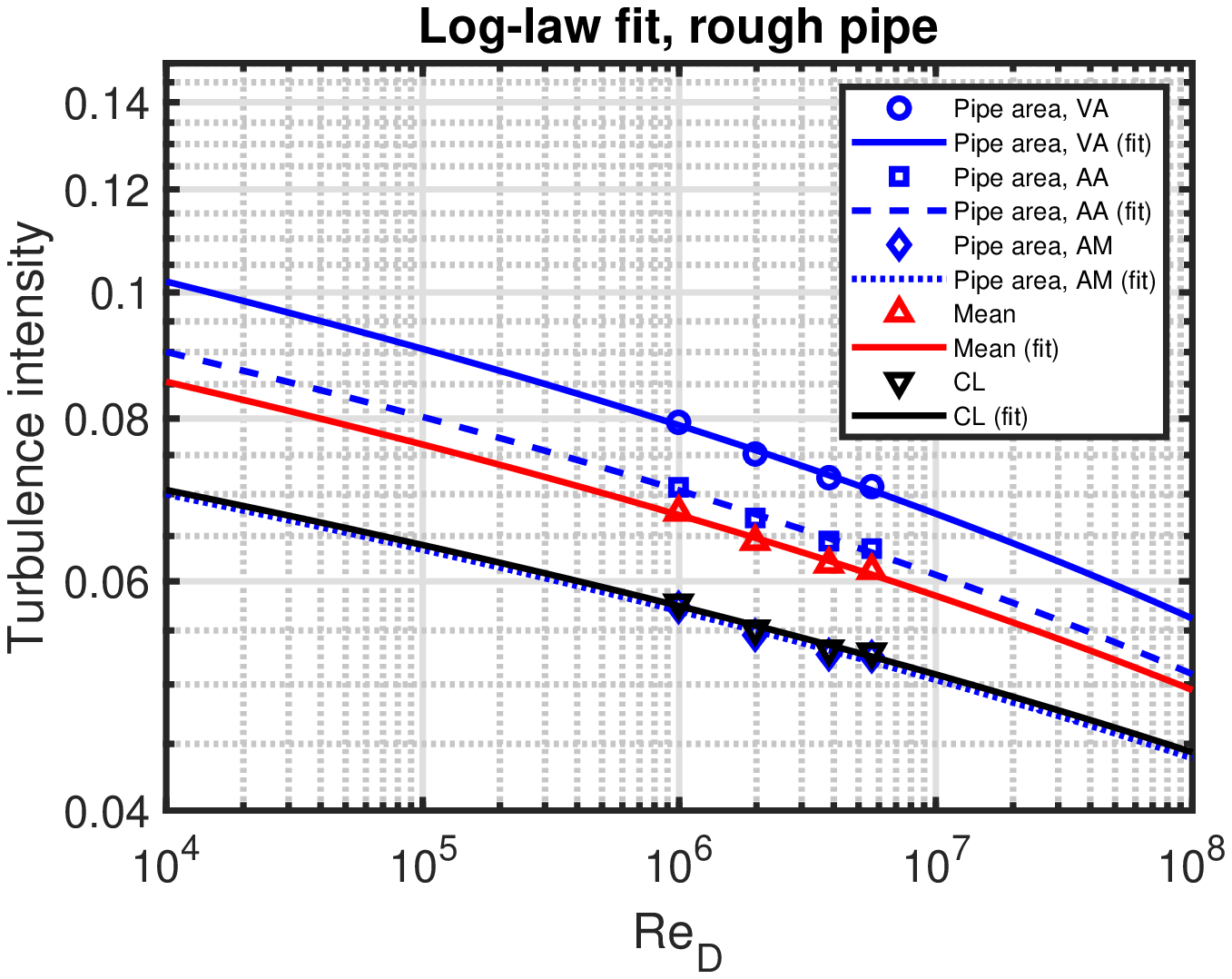}
\caption{Rough pipe turbulence intensity as a function of $Re_{\rm D}$, left: Power-law fit, right: Log-law fit.}
\label{fig:log_I_smooth_rough}
\end{figure}

The results are presented in Figures \ref{fig:I_smooth_rough} and \ref{fig:log_I_smooth_rough} and Tables \ref{tab:smooth} to \ref{tab:log-rough}.

We do not discuss the quality of fits to the rough pipe, since there are only 4 measurements for a single $k_s$. Thus, these values are provided as a reference.

For the smooth pipe, the power-law fits perform slightly better than the log-law fits, except for the CL definition. The best fit is using the power-law fit to the AA definition of the TI:

\begin{equation}
\label{eq:best_ti}
I_{\rm Pipe~area,~AA} = 0.3173 \times Re_{\rm D}^{-0.1095},
\end{equation}

\noindent which is the same as Equation (5) in \cite{basse_a}.

\begin{table}
\centering
\caption{Power-law fit constants, smooth pipe.}
%% \tablesize{} %% You can specify the fontsize here, e.g.  \tablesize{\footnotesize}. If commented out \small will be used.
{\begin{tabular}{cccc}
\hline
\textbf{TI definition}	& \textbf{a}	& \textbf{b} & \textbf{RMSD}\\
\hline
$I_{\rm Pipe~area,~AM}$		& 0.2274			& -0.1004 & 4.0563 $\times$ 10$^{-4}$\\
$I_{\rm Pipe~area,~AA}$		& 0.3173			& -0.1095 & 3.5932 $\times$ 10$^{-4}$\\
$I_{\rm Pipe~area,~VA}$		& 0.3758			& -0.1134 & 3.6210 $\times$ 10$^{-4}$\\
$I_m$                       & 0.2657			& -0.1000 & 5.2031 $\times$ 10$^{-4}$\\
$I_{\rm CL}$                & 0.1811			& -0.0837 & 6.9690 $\times$ 10$^{-4}$\\
\hline
\end{tabular}}
\label{tab:smooth}
\end{table}

\begin{table}
\centering
\caption{Log-law fit constants, smooth pipe.}
%% \tablesize{} %% You can specify the fontsize here, e.g.  \tablesize{\footnotesize}. If commented out \small will be used.
{\begin{tabular}{cccc}
\hline
\textbf{TI definition}	& \textbf{c}	& \textbf{d} & \textbf{RMSD}\\
\hline
$I_{\rm Pipe~area,~AM}$		& -0.0059			& 0.1391 & 6.7748 $\times$ 10$^{-4}$\\
$I_{\rm Pipe~area,~AA}$		& -0.0080			& 0.1808 & 8.8173 $\times$ 10$^{-4}$\\
$I_{\rm Pipe~area,~VA}$		& -0.0093			& 0.2074 & 1.1018 $\times$ 10$^{-3}$\\
$I_m$                       & -0.0069			& 0.1634 & 5.8899 $\times$ 10$^{-4}$\\
$I_{\rm CL}$                & -0.0049			& 0.1257 & 5.4179 $\times$ 10$^{-4}$\\
\hline
\end{tabular}}
\label{tab:log-smooth}
\end{table}

\begin{table}
\centering
\caption{Power-law fit constants, rough pipe.}
%% \tablesize{} %% You can specify the fontsize here, e.g.  \tablesize{\footnotesize}. If commented out \small will be used.
{\begin{tabular}{cccc}
\hline
\textbf{TI definition}	& \textbf{a}	& \textbf{b}  & \textbf{RMSD}\\
\hline
$I_{\rm Pipe~area,~AM}$		& 0.1172			& -0.0522 & 4.0830 $\times$ 10$^{-4}$\\
$I_{\rm Pipe~area,~AA}$		& 0.1702			& -0.0638 & 3.5784 $\times$ 10$^{-4}$\\
$I_{\rm Pipe~area,~VA}$		& 0.1989			& -0.0667 & 3.7697 $\times$ 10$^{-4}$\\
$I_m$                       & 0.1568			& -0.0610 & 3.4902 $\times$ 10$^{-4}$\\
$I_{\rm CL}$                & 0.1177			& -0.0519 & 3.4317 $\times$ 10$^{-4}$\\
\hline
\end{tabular}}
\label{tab:rough}
\end{table}

\begin{table}
\centering
\caption{Log-law fit constants, rough pipe.}
%% \tablesize{} %% You can specify the fontsize here, e.g.  \tablesize{\footnotesize}. If commented out \small will be used.
{\begin{tabular}{cccc}
\hline
\textbf{TI definition}	& \textbf{c}	& \textbf{d}  & \textbf{RMSD}\\
\hline
$I_{\rm Pipe~area,~AM}$		& -0.0028			& 0.0960 & 4.3111 $\times$ 10$^{-4}$\\
$I_{\rm Pipe~area,~AA}$		& -0.0042			& 0.1291 & 4.0028 $\times$ 10$^{-4}$\\
$I_{\rm Pipe~area,~VA}$		& -0.0050			& 0.1477 & 4.2938 $\times$ 10$^{-4}$\\
$I_m$                       & -0.0039			& 0.1213 & 3.8575 $\times$ 10$^{-4}$\\
$I_{\rm CL}$                & -0.0028			& 0.0967 & 3.6542 $\times$ 10$^{-4}$\\
\hline
\end{tabular}}
\label{tab:log-rough}
\end{table}

Instead of $Re_{\rm D}$, one can also express the TI fits using the friction Reynolds number \cite{hultmark_a}:

\begin{equation}
Re_{\tau}=\frac{R v_{\tau}}{\nu_{\rm kin}} = \frac{v_{\tau}}{2 v_m} \times Re_{\rm D}
\end{equation}

\begin{figure}
%\vspace{0.5cm}
\centering
\includegraphics[width=10cm]{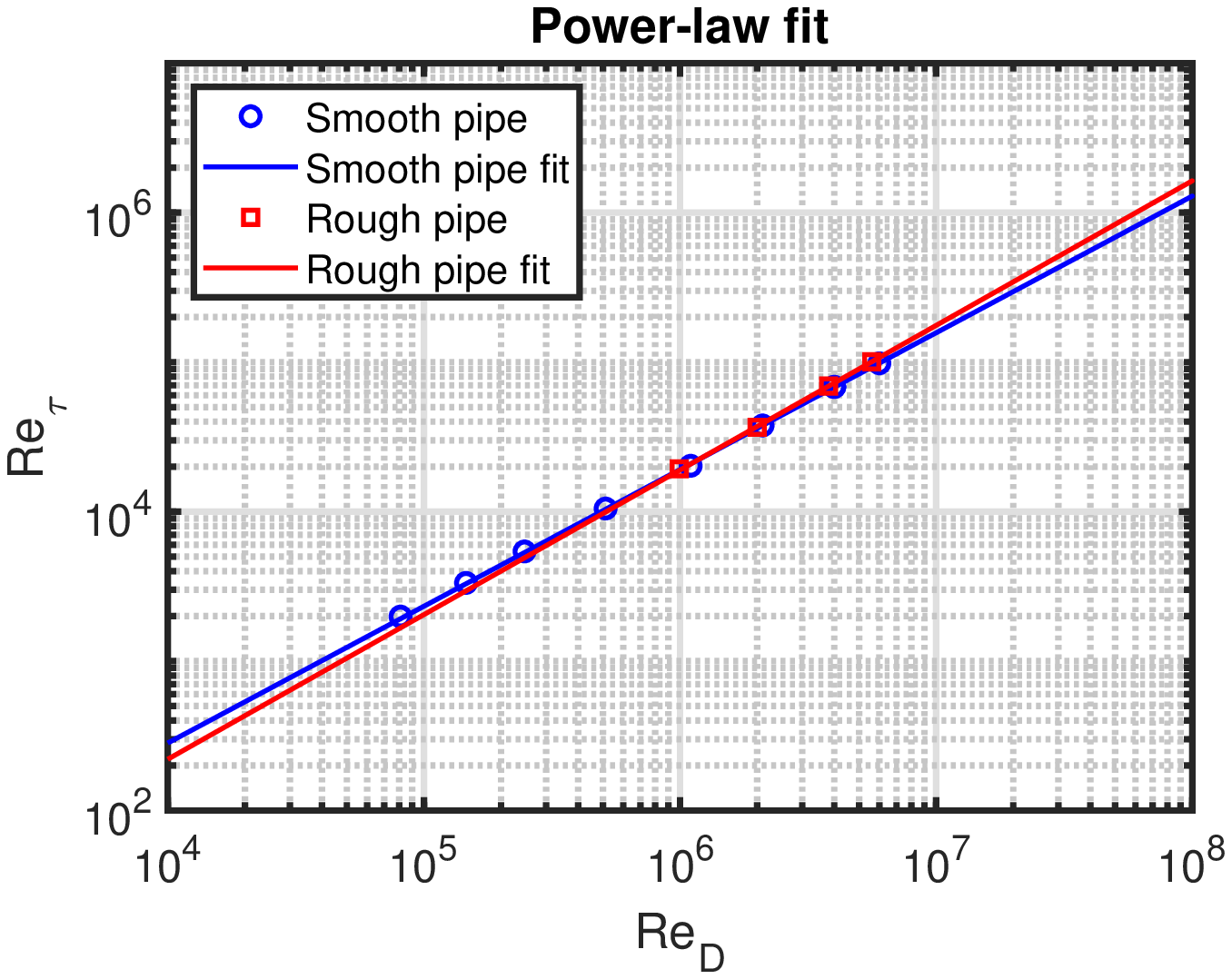}
\caption{Relationship between $Re_{\rm D}$ and $Re_{\tau}$.}
\label{fig:AA_Re_D_Re_tau}
\end{figure}

The relationship between $Re_{\rm D}$ and $Re_{\tau}$ can fitted using Equation (\ref{eq:pow-fit}) where $Q=Re_{\tau}$ and $x=Re_{\rm D}$, see Figure \ref{fig:AA_Re_D_Re_tau} and Table \ref{tab:Re}. A log-law fit was also performed but resulted in a bad fit, i.e. a RMSD which was between one and two orders of magnitude larger than for the power-law fit. As mentioned above, the rough pipe fit is only provided as a reference. For channel flow, it has been found that $a=0.09$ and $b=0.88$, see Figure 7.11 in \cite{pope_a} and associated text.

We choose to focus on the bulk Reynolds number since it is possible to determine for applications where the friction velocity is unknown.

\begin{table}
\centering
\caption{Bulk and friction Reynolds number fits.}
%% \tablesize{} %% You can specify the fontsize here, e.g.  \tablesize{\footnotesize}. If commented out \small will be used.
{\begin{tabular}{ccc}
\hline
\textbf{Case}	& \textbf{a}	& \textbf{b}\\
\hline
Smooth		    & 0.0621		& 0.9148\\
Rough           & 0.0297		& 0.9675\\
\hline
\end{tabular}}
\label{tab:Re}
\end{table}

\section{Discussion}
\label{sec:disc}

\subsection{Friction Factor}
\label{subsec:ff}

In \cite{mckeon_a}, the following expression for the smooth pipe friction factor has been derived based on Princeton Superpipe measurements:

\begin{equation}
\frac{1}{\sqrt{\lambda_{\rm Smooth}}} = 1.930 \log_{10} \left( Re_{\rm D} \sqrt{\lambda_{\rm Smooth}}\right) - 0.537
\label{eq:ff_mckeon}
\end{equation}

A corresponding rough pipe friction factor has been proposed in \cite{colebrook_a}:

\begin{equation}
\frac{1}{\sqrt{\lambda_{\rm Rough}}} = -2 \log_{10} \left( \frac{k_s}{3.7D} + \frac{2.51}{Re_{\rm D} \sqrt{\lambda_{\rm Rough}}}\right)
\label{eq:ff_colebrook}
\end{equation}

The friction factor can also be expressed using the friction velocity and the mean velocity, see Equation (1.1) in \cite{mckeon_a}:

\begin{equation}
 \lambda = 8 \times \frac{v_{\tau}^2}{v_m^2}
\label{eq:fric_mani}
\end{equation}

\noindent or:

\begin{equation}
\frac{v_{\tau}}{v_m} = \sqrt{\frac{\lambda}{8}}
\label{eq:ratio-lambda}
\end{equation}

\begin{figure}
\centering
\includegraphics[width=7cm]{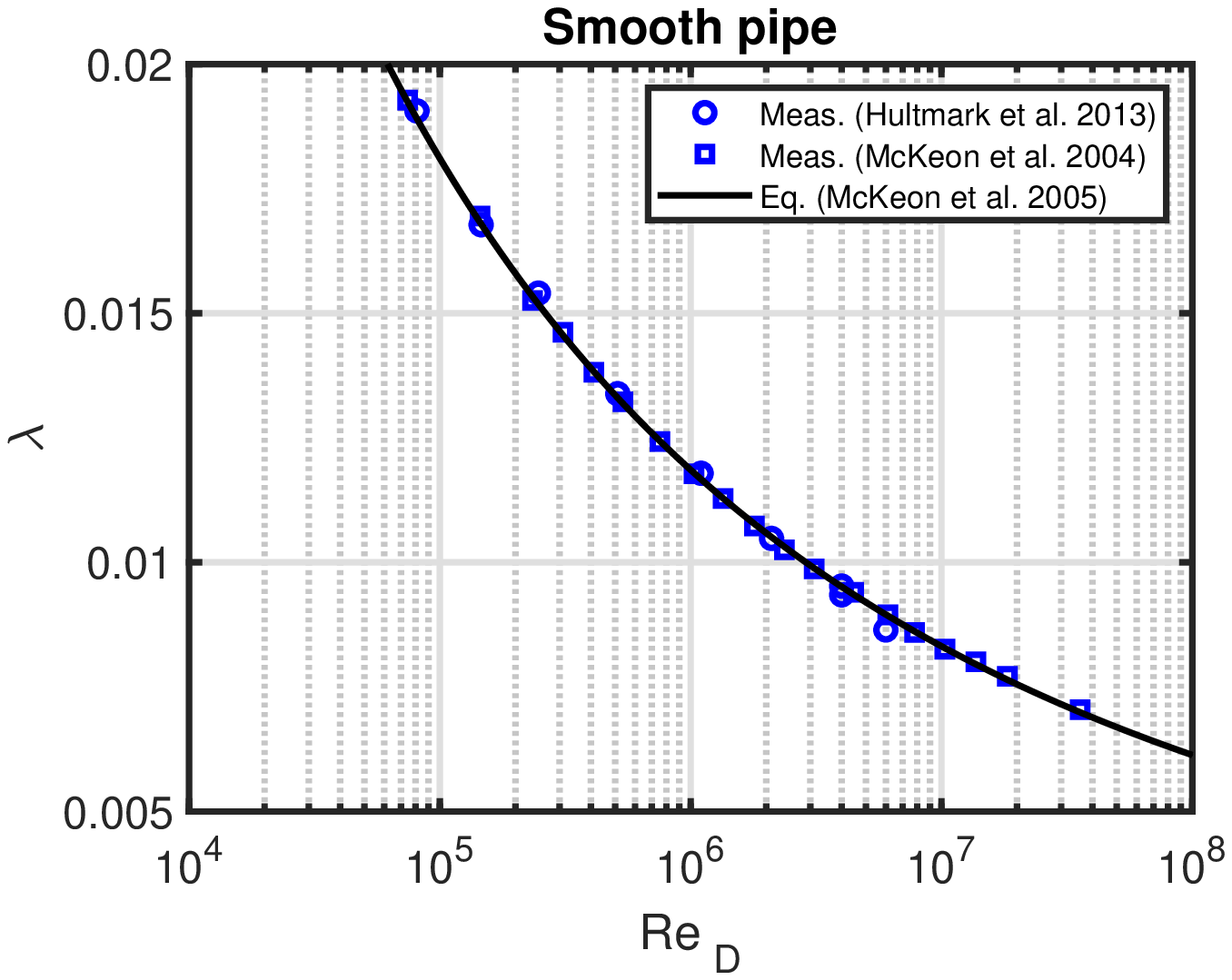}
\includegraphics[width=7cm]{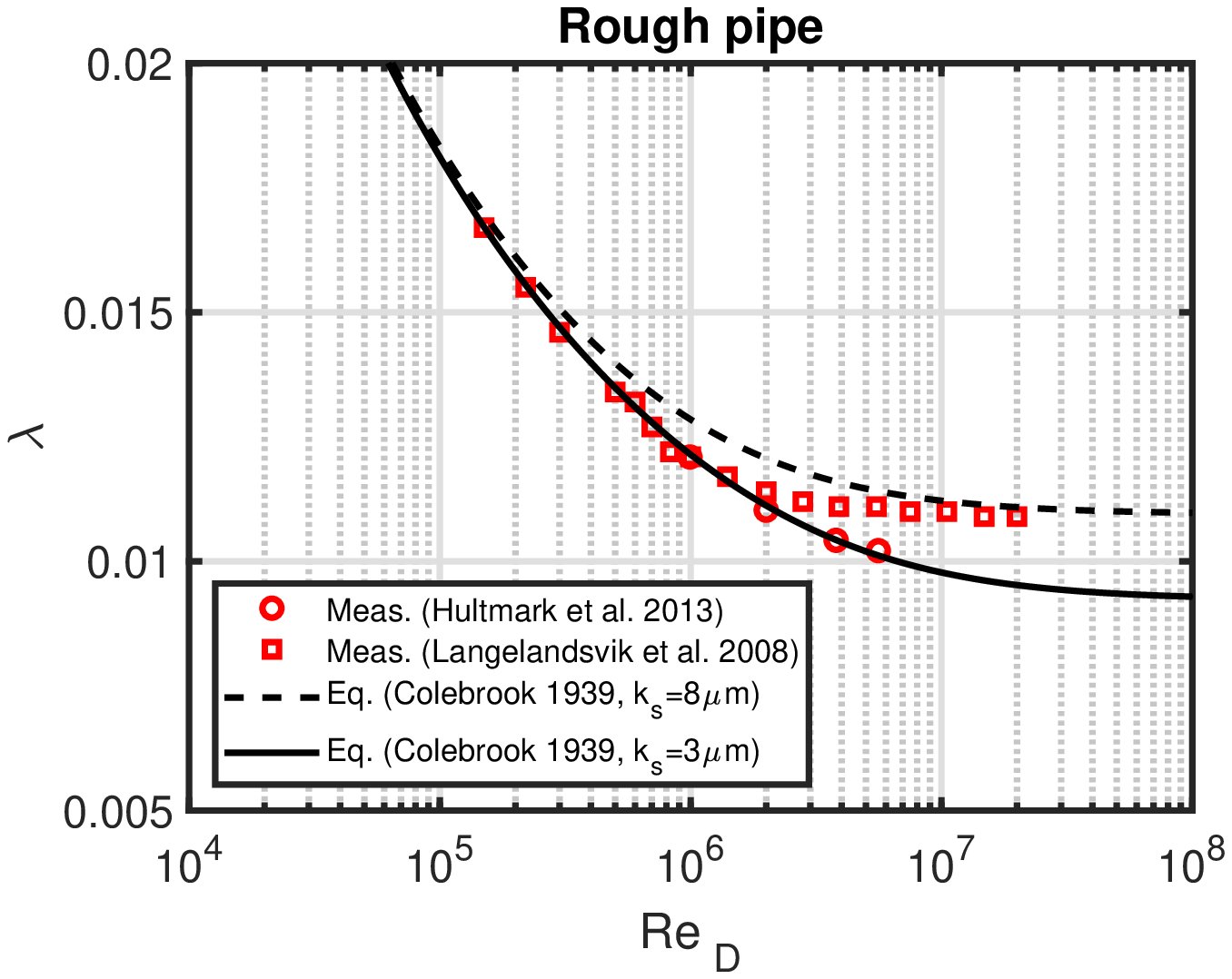}
\caption{Friction factors, left: Smooth pipe, right: Rough pipe.}
\label{fig:fric_fac}
\end{figure}

The equations for the smooth- and rough-wall pipe flow friction factors are shown in Figure \ref{fig:fric_fac} along with Princeton Superpipe measurements.

We have included additional smooth pipe measurements \cite{mckeon_b,smits_b}. Both sets agree with Equation (\ref{eq:ff_mckeon}).

Additional rough pipe measurements can be found in Table 2 (and Figure 3) in \cite{langelandsvik_a}. Here, it was found that $k_s = 8~\mu$m. For our main data set \cite{hultmark_a,smits_a}, $k_s = 8~\mu$m does not match the measurements; instead we get $k_s=3~\mu$m for a fit to those measurements using Equation (\ref{eq:ff_colebrook}). It is the same pipe; the reason for the discrepancy is not clear \cite{hultmark_b}. However, the difference is within the experimental uncertainty of 5\% stated in \cite{langelandsvik_a}.

For the rough pipe friction factor, we use $k_s=3~\mu$m in the remainder of this paper.

\subsection{Turbulence Intensity Aspects}

\subsubsection{Importance for Flow}

The TI is an important quantity for many physical phenomena \cite{schlichting_a}, e.g.:
\begin{itemize}
  \item The critical Reynolds number for the drag of a sphere \cite{zarin_a}
  \item The laminar-turbulent transition \cite{fransson_a}
  \item Development of the turbulent boundary layer \cite{hollingsworth_a}
  \item The position of flow separation \cite{scheichl_a}
  \item Heat transfer \cite{ahn_a}
  \item Wind farms \cite{hansen_a}
  \item Wind tunnels \cite{calautit_a}
\end{itemize}

We mention these examples to illustrate the importance of the TI for real world applications.

\subsubsection{Scaling with the Friction Factor}

The wall-normal \cite{orlandi_a} and streamwise (this paper and \cite{basse_a}) Reynolds stress have both been shown to be linked to the friction factor $\lambda = 4C_f$, where $C_f$ is the skin friction coefficient. These observations can be interpreted as manifestations of the FIK identity \cite{fukagata_a}, where an equation for $C_f$ is derived based on the streamwise momentum equation: $C_f$ is proportional to the integral over the Reynolds shear stress weighted by the quadratic distance from the pipe axis.

An alternative formulation for $C_f$ based on the streamwise kinetic energy is derived in \cite{renard_a}: Here, $C_f$ is proportional to the integral over the Reynolds shear stress multiplied by the streamwise mean velocity gradient weighted by the distance from the pipe axis. It is concluded that the logarithmic region dominates friction generation for high Reynolds number flow. The dominance of the logarithmic region has been confirmed in \cite{giovanetti_a}.

\begin{figure}
%\vspace{0.5cm}
\centering
\includegraphics[width=7cm]{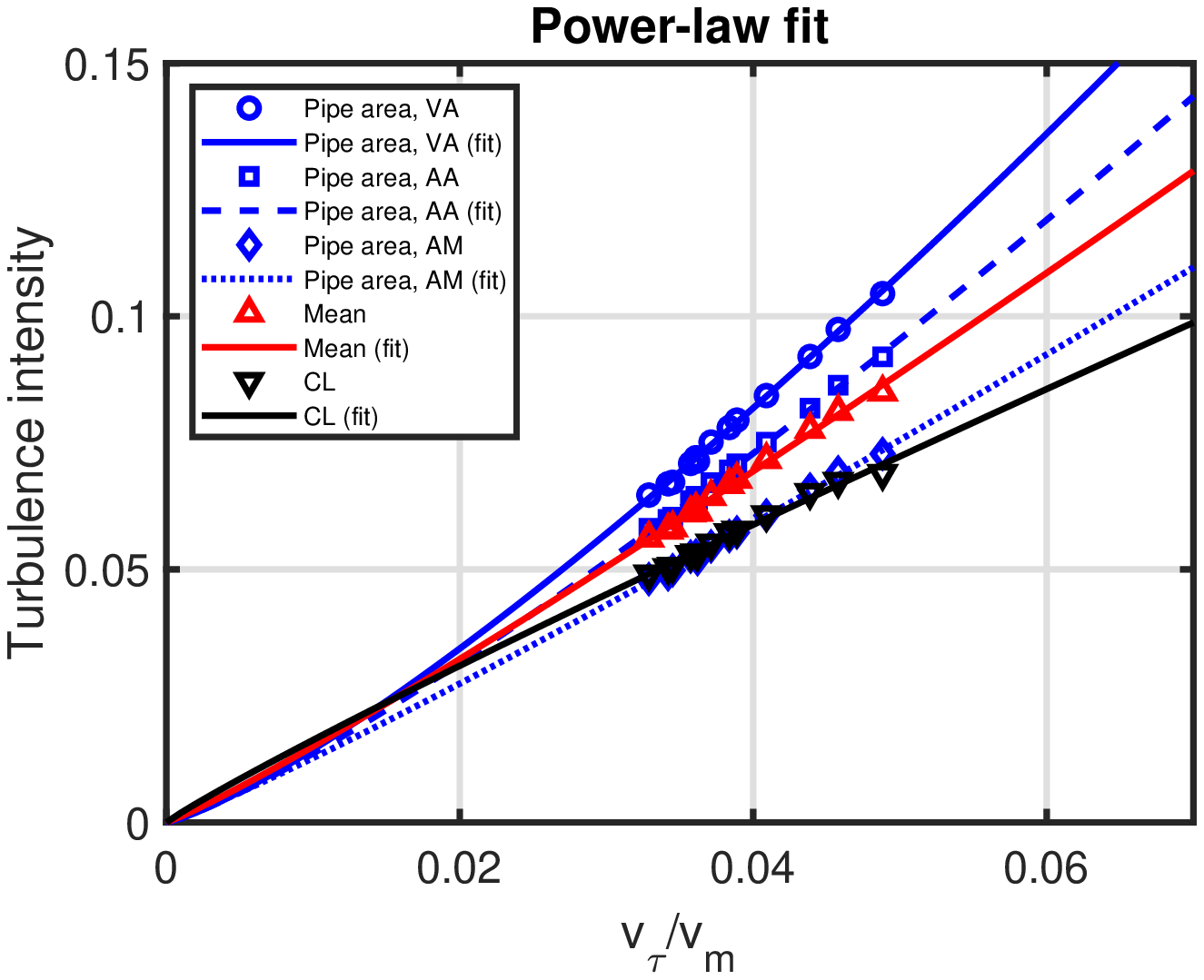}
\includegraphics[width=7cm]{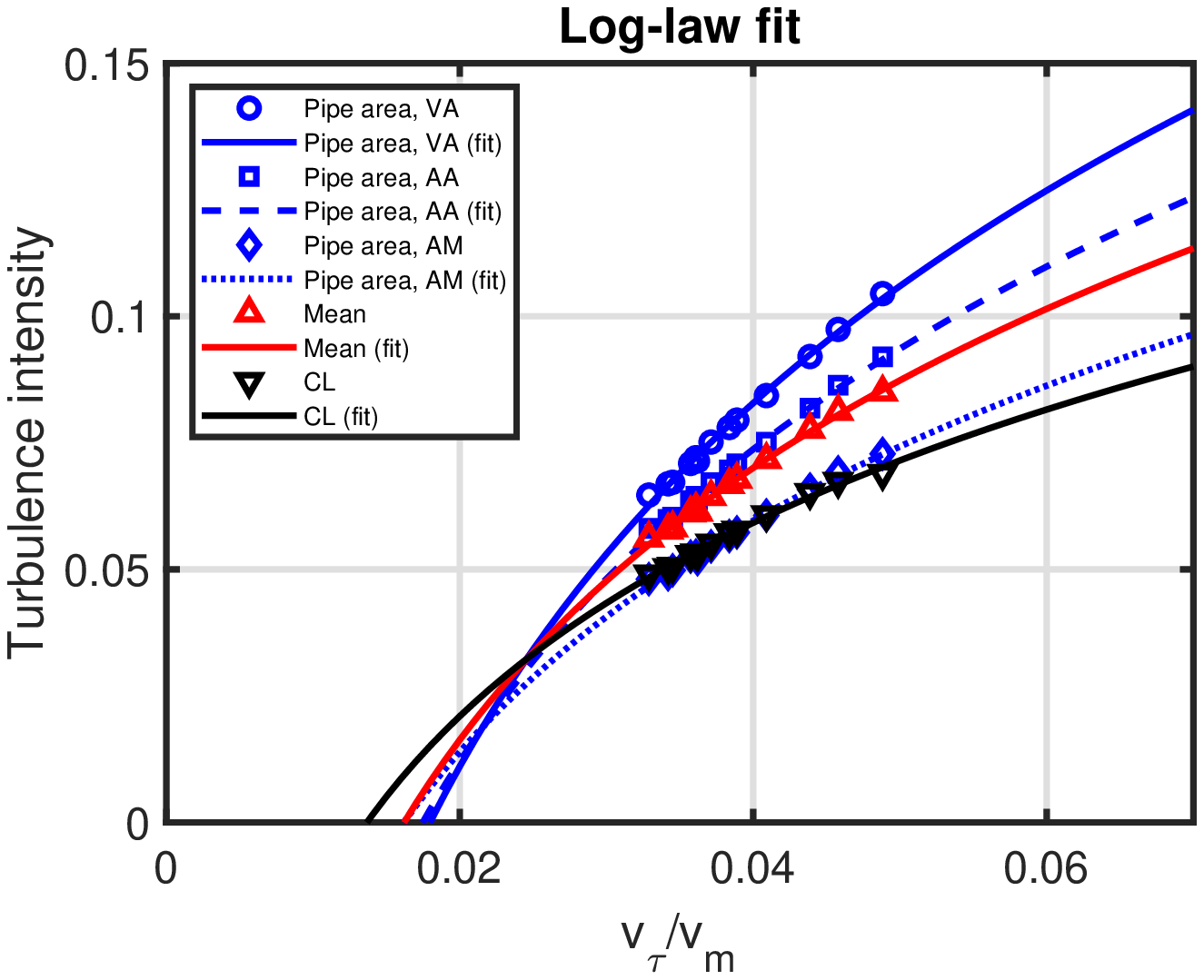}
\caption{TI as a function of $v_{\tau}/v_m$.}
\label{fig:TI_vs_fric_vm}
\end{figure}

The seeming equivalence between Reynolds stress (shear or normal) and the friction factor leads us to propose that the TI scales with $v_{\tau}/v_m$. Therefore we fit to Equations (\ref{eq:pow-fit}) and (\ref{eq:log-fit}) using $Q=I$ and $x=v_{\tau}/v_m$, see Figure \ref{fig:TI_vs_fric_vm} and Tables \ref{tab:TI-smooth} and \ref{tab:TI-log-smooth}.

In this case, the power-law fits are best for the local velocity definitions and the log-law fits are best for the reference velocity definitions. Overall, the best fit is the power-law fit using the AM definition of the TI:

\begin{equation}
\label{eq:best_fric}
I_{\rm Pipe~area,~AM} = 0.6577 \times \lambda^{0.5531},
\end{equation}

\noindent which is a modification of Equation (14) in \cite{basse_a}.

\begin{table}
\centering
\caption{Power-law fit constants.}
%% \tablesize{} %% You can specify the fontsize here, e.g.  \tablesize{\footnotesize}. If commented out \small will be used.
{\begin{tabular}{cccc}
\hline
\textbf{TI definition}	& \textbf{a}	& \textbf{b} & \textbf{RMSD}\\
\hline
$I_{\rm Pipe~area,~AM}$		& 2.0776			& 1.1062 & 4.7133 $\times$ 10$^{-4}$\\
$I_{\rm Pipe~area,~AA}$		& 3.5702			& 1.2088 & 5.0224 $\times$ 10$^{-4}$\\
$I_{\rm Pipe~area,~VA}$		& 4.6211			& 1.2530 & 4.9536 $\times$ 10$^{-4}$\\
$I_m$                       & 2.4238			& 1.1039 & 6.8412 $\times$ 10$^{-4}$\\
$I_{\rm CL}$                & 1.1586			& 0.9260 & 7.5876 $\times$ 10$^{-4}$\\
\hline
\end{tabular}}
\label{tab:TI-smooth}
\end{table}

\begin{table}
\centering
\caption{Log-law fit constants.}
%% \tablesize{} %% You can specify the fontsize here, e.g.  \tablesize{\footnotesize}. If commented out \small will be used.
{\begin{tabular}{cccc}
\hline
\textbf{TI definition}	& \textbf{c}	& \textbf{d} & \textbf{RMSD}\\
\hline
$I_{\rm Pipe~area,~AM}$		& 0.0658			& 0.2715 & 5.5459 $\times$ 10$^{-4}$\\
$I_{\rm Pipe~area,~AA}$		& 0.0890			& 0.3602 & 6.6955 $\times$ 10$^{-4}$\\
$I_{\rm Pipe~area,~VA}$		& 0.1036			& 0.4162 & 8.0775 $\times$ 10$^{-4}$\\
$I_m$                       & 0.0775			& 0.3195 & 5.5182 $\times$ 10$^{-4}$\\
$I_{\rm CL}$                & 0.0551			& 0.2367 & 5.8536 $\times$ 10$^{-4}$\\
\hline
\end{tabular}}
\label{tab:TI-log-smooth}
\end{table}

Combining Equation (\ref{eq:ratio-lambda}) with Equations (\ref{eq:pow-fit}) and (\ref{eq:log-fit}) leads to:

\begin{equation}
I_{\rm Power-law~fit} (\lambda) = a \times \left( \frac{\lambda}{8} \right)^{b/2}
\label{eq:I-pow}
\end{equation}

\begin{equation}
I_{\rm Log-law~fit} (\lambda) = \frac{c}{2} \times \ln \left( \frac{\lambda}{8} \right) + d
\label{eq:I-log}
\end{equation}

\begin{figure}
\centering
\includegraphics[width=10cm]{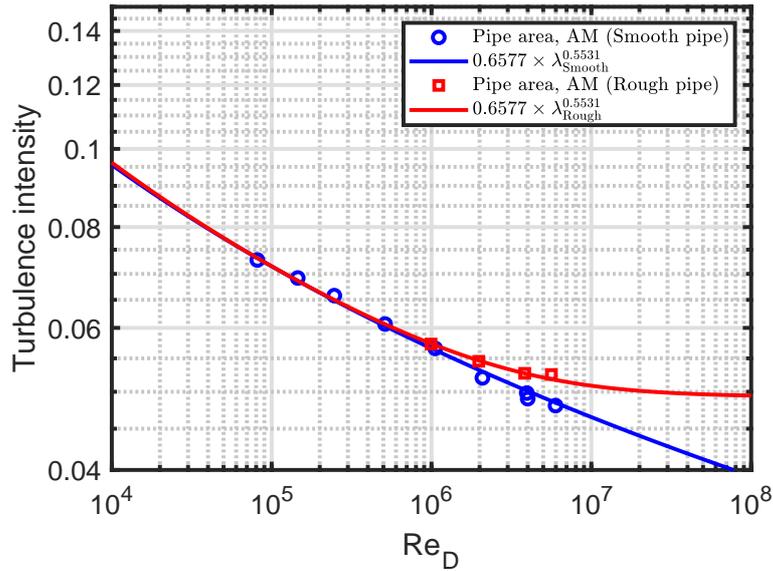}
\caption{AM definition of TI as a function of $Re_{\rm D}$ for smooth- and rough-wall pipe flow.}
\label{fig:comp_fric_ti}
\end{figure}

The predicted TI for the best case (power-law AM) is shown in Figure \ref{fig:comp_fric_ti}; one reason for the better match to measurements compared to Figure 9 in \cite{basse_a} is that we use $k_s=3~\mu$m instead of $k_s=8~\mu$m for the rough pipe. The correspondence between the TI and the friction factor means that the TI will approach a constant value for rough-wall pipe flow at large $Re_{\rm D}$ (fully rough regime). It also means that a larger $k_s$ leads to a higher TI.

\subsubsection{CFD Definition}

Let us now consider a typical CFD turbulence model, the standard $k-\varepsilon$ model \cite{versteeg_a}. Here, $k$ is the turbulent kinetic energy (TKE) per unit mass and $\varepsilon$ is the rate of dissipation of TKE per unit mass.

As an example of a boundary condition, the user provides the TI ($I_{\rm user}$) and the turbulent viscosity ratio $\mu_t/\mu$, where $\mu_t$ is the dynamic turbulent viscosity and $\mu$ is the dynamic viscosity. For a defined reference velocity $v_{\rm ref}$, $k$ can then be calculated as:

\begin{equation}
%\label{}
k=\frac{3}{2} \left( v_{\rm ref} I_{\rm user} \right)^2
\end{equation}

As the next step, $\varepsilon$ is defined as:

\begin{equation}
%\label{}
\varepsilon=\frac{\rho C_{\mu} k^2}{\left( \mu_t/\mu \right) \mu},
\end{equation}

\noindent where $\rho$ is density and $C_{\mu}=0.09$.

An example of default CFD settings is $I_{\rm user}=0.01$ (1\%) and $\mu_t/\mu=10$.

The output from a CFD simulation is the total TKE, not the individual components. If we assume that the turbulence is isotropic, the streamwise TI we are treating in this paper is proportional to the square root of the TKE:

\begin{equation}
%\label{}
v_{\rm RMS} = \sqrt{\frac{2}{3}k}
\end{equation}

\subsubsection{Proposed Procedure for CFD and an Example}
\label{subsubsec:prop}

A standard definition of TI for CFD is to use the free-stream velocity as the reference velocity, i.e $I_{\rm CL}$ for pipe flow. For $I_{\rm CL}$, we use the log-law version since this has the smallest RMSD, see Tables \ref{tab:TI-smooth} and \ref{tab:TI-log-smooth}:

\begin{equation}
\label{eq:best_CL}
I_{\rm CL} = 0.0276 \times \ln (\lambda) + 0.1794
\end{equation}

We note that this scaling is based on the Princeton Superpipe measurements; for industrial applications, the TI may be much higher, so our scaling should be considered as a lower limit.

A procedure to calculate the TI e.g. for use in CFD is:
\begin{enumerate}
  \item Define $Re_{\rm D}$ (and $k_s$ for a rough pipe)
  \item Calculate the friction factor: Equation (\ref{eq:ff_mckeon}) for a smooth pipe and Equation (\ref{eq:ff_colebrook}) for a rough pipe
  \item Use Equation (\ref{eq:best_CL}) to calculate the TI
\end{enumerate}

As a concrete example, we consider incompressible (water) flow through a 130 mm diameter pipe. CFD boundary conditions can be velocity inlet and pressure outlet. The steps are:
\begin{enumerate}
  \item Define the mean velocity, we use 10 m/s
  \item Calculate $Re_{\rm D}$ = $1.3 \times 10^6$
  \item Use Equation (\ref{eq:ff_mckeon}) to calculate $\lambda_{\rm Smooth}=0.0114$
  \item Use Equation (\ref{eq:best_CL}) to calculate $I_{\rm CL}=0.0561$
\end{enumerate}

For this example, we conclude that the minimum TI is 5.6\%.

\subsubsection{Open Questions}

It remains an open question to what extent the quality of the fits (RMSD) impacts the outcome of a CFD simulation. For the TI definitions used, the RMSD varies less than a factor of two for the fits of TI as a function of $v_{\tau}/v_m$, see Tables \ref{tab:TI-smooth} and \ref{tab:TI-log-smooth}. For CFD, a single TI definition is used, so it is not possible to switch TI models in CFD and compare simulations to measurements.

To continue our measurement-based research, we would need measurements of all velocity components, i.e. wall-normal and spanwise in addition to the available streamwise measurements.

As mentioned earlier \cite{russo_a}, it would also be interesting to have measurements for higher Mach numbers, where compressibility will play a larger role.

In addition to pipe flow, other canonical flows, such as zero-pressure gradient flows \cite{marusic_b} might be suitable for analysis similar to what we have presented.

\section{Conclusions}
\label{sec:conc}

We have used Princeton Superpipe measurements of smooth- and rough-wall pipe flow \cite{hultmark_a,smits_a} to study the properties of various TI definitions. Scaling of TI with $Re_{\rm D}$ is provided for the definitions. For scaling purposes, we recommend the AA definition and a power-law fit: Equation (\ref{eq:best_ti}). The TI also scales with $v_{\tau}/v_m$, where the best result is obtained with a power-law fit and the AM definition. This fit implies that the turbulence level scales with the friction factor: Equation (\ref{eq:best_fric}).

Scaling of TI with $Re_{\rm D}$ and $v_{\tau}/v_m$ was done using both power-law and log-law fits.

A proposed procedure to calculate the TI for e.g. CFD is provided and exemplified in Section \ref{subsubsec:prop}.

%%%%%%%%%%%%%%%%%%%%%%%%%%%%%%%%%%%%%%%%%%
\acknowledgments{We thank Professor Alexander J. Smits for making the Princeton Superpipe data publicly available.}

%%%%%%%%%%%%%%%%%%%%%%%%%%%%%%%%%%%%%%%%%%
\conflictsofinterest{The authors declare no conflict of interest.}

%%%%%%%%%%%%%%%%%%%%%%%%%%%%%%%%%%%%%%%%%%
%% optional

%=====================================
% References, variant A: internal bibliography
%=====================================
\reftitle{References}

% The following MDPI journals use author-date citation: Arts, Econometrics, Economies, Genealogy, Humanities, IJFS, JRFM, Laws, Religions, Risks, Social Sciences. For those journals, please follow the formatting guidelines on http://www.mdpi.com/authors/references
% To cite two works by the same author: \citeauthor{ref-journal-1a} (\citeyear{ref-journal-1a}, \citeyear{ref-journal-1b}). This produces: Whittaker (1967, 1975)
% To cite two works by the same author with specific pages: \citeauthor{ref-journal-3a} (\citeyear{ref-journal-3a}, p. 328; \citeyear{ref-journal-3b}, p.475). This produces: Wong (1999, p. 328; 2000, p. 475)

%=====================================
% References, variant B: external bibliography
%=====================================
%\externalbibliography{yes}
%\bibliography{your_external_BibTeX_file}

\begin{thebibliography}{999}
\bibitem{ansys_a} ANSYS Fluent User's Guide, Release 19.0, Section 6.3.2.1.3, 2018.
\bibitem{russo_a} Russo, F.; Basse, N.T. Scaling of turbulence intensity for low-speed flow in smooth pipes. \emph{Flow Meas. Instrum}. \textbf{2016}, \emph{52}, 101--114.
\bibitem{basse_a} Basse, N.T. Turbulence intensity and the friction factor for smooth- and rough-wall pipe flow. \emph{Fluids} \textbf{2017}, \emph{2}, 30.
\bibitem{hultmark_a} Hultmark, M.; Vallikivi, M.; Bailey, S.C.C.; Smits, A.J. Logarithmic scaling of turbulence in smooth- and rough-wall pipe flow. \emph{J. Fluid Mech.} \textbf{2013}, \emph{728}, 376--395.
\bibitem{smits_a} Princeton Superpipe. Available online: \newline \texttt{https://smits.princeton.edu/superpipe-turbulence-data} (accessed on 5 July 2019).
\bibitem{willert_a} Willert, C.E.; et al. Near-wall statistics of a turbulent pipe flow at shear Reynolds numbers up to 40 000. \emph{J. Fluid Mech.} \textbf{2017}, \emph{826}, R5.
\bibitem{schultz_a} Schultz, M.P.; Flack, K.A. The rough-wall turbulent boundary layer from the hydraulically smooth to the fully rough regime. \emph{J. Fluid Mech.} \textbf{2007}, \emph{580}, 381--405.
\bibitem{flack_a} Flack, K.A.; Schultz, M.P. Roughness effects on wall-bounded turbulent flows. \emph{Phys. Fluids}. \textbf{2014}, \emph{26}, 101305.
\bibitem{marusic_a} Marusic, I.; Baars, W.J.; Hutchins, N. Scaling of the streamwise turbulence intensity in the context of inner-outer
interactions in wall turbulence. \emph{Phys. Rev. Fluids} \textbf{2017}, \emph{2}, 100502.
\bibitem{laufer_a} Laufer, J. Investigation of turbulent flow in a two-dimensional channel. \emph{NACA-TR-1053} \textbf{1951}.
\bibitem{laufer_b} Laufer, J. The structure of turbulence in fully developed pipe flow. \emph{NACA-TR-1174} \textbf{1954}.
\bibitem{touponce_a} Touponce, W.F. \emph{Frank Herbert}; Twayne Publishers: Boston, USA, 1988.
\bibitem{herbert_a} Frank Herbert's Classic Dune: \newline \texttt{http://www.dunenovels.com/frank-herbert-classic-dune} (accessed on 5 July 2019).
\bibitem{marusic_b} Marusic, I.; Kunkel, G.J. Streamwise turbulence intensity formulation for flat-plate boundary layers. \emph{Phys. Fluids} \textbf{2003}, \emph{15}, 2461--2464.
\bibitem{alfredsson_a} Alfredsson, P.H.; \"Orl\"u, R.; Segalini, A. A new formulation for the streamwise turbulence intensity distribution in wall-bounded turbulent flows. \emph{Eur. J. Mech. B Fluids} \textbf{2012}, \emph{36}, 167--175.
\bibitem{monkewitz_a} Monkewitz, P.A.; Nagib, H.M. Large-Reynolds number asymptotics of the streamwise normal stress in zero-pressure-gradient turbulent boundary layers. \emph{J. Fluid Mech.} \textbf{2015}, \emph{783}, 474--503.
\bibitem{orlandi_a} Orlandi, P. The importance of wall-normal Reynolds stress in turbulent rough channel flows. \emph{Phys. Fluids} \textbf{2013}, \emph{25}, 110813.
\bibitem{fukagata_a} Fukagata, K.; Iwamoto, K.; Kasagi, N. Contribution of Reynolds stress distribution to the skin friction in wall-bounded flows. \emph{Phys. Fluids}. \textbf{2002}, \emph{14}, L73--L76.
\bibitem{gersten_a} Gersten, K. Fully developed turbulent pipe flow, in: Merzkirch, W. (Ed.) \emph{Fluid Mechanics of Flow Metering}; Springer: Berlin, Germany, 2005.
\bibitem{smits_b} Princeton Superpipe. Available online: \newline \texttt{https://smits.princeton.edu/mckeon} (accessed on 5 July 2019).
\bibitem{barenblatt_a} Barenblatt, G.I. Scaling laws for fully developed turbulent shear flows. Part 1. Basic hypotheses and analysis. \emph{J.~Fluid Mech.} \textbf{1993}, \emph{248}, 513--520.
\bibitem{zagarola_a} Zagarola, M.V.; Perry, A.E.; Smits, A.J. Log laws or power laws: The scaling in the overlap region. \emph{Phys. Fluids} \textbf{1997}, \emph{9}, 2094--2100.
\bibitem{zagarola_b} Zagarola, M.V.; Smits, A.J. Mean-flow scaling of turbulent pipe flow. \emph{J.~Fluid Mech.} \textbf{1998}, \emph{373}, 33--79.
\bibitem{langelandsvik_a} Langelandsvik, L.I.; Kunkel, G.J.; Smits, A.J. Flow in a commercial steel pipe. \emph{J.~Fluid Mech.} \textbf{2008}, \emph{595}, 323--339.
\bibitem{pope_a} Pope, S.B. \emph{Turbulent flows}; Cambridge University Press: Cambridge, UK, 2000.
\bibitem{mckeon_a} McKeon, B.J.; Zagarola, M.V.; Smits, A.J. A new friction factor relationship for fully developed pipe flow. \emph{J.~Fluid Mech.} \textbf{2005}, \emph{538}, 429--443.
\bibitem{colebrook_a} Colebrook, C.F. Turbulent flow in pipes, with particular reference to the transition region between the smooth and rough pipe laws. \emph{J.~Inst.~Civil~Engrs.} \textbf{1939}, \emph{11}, 133--156.
\bibitem{mckeon_b} McKeon, B.J.; Li, J.; Jiang, W.; Morrison, J.F; Smits, A.J. Further observations on the mean velocity
distribution in fully developed pipe flow. \emph{J.~Fluid Mech.} \textbf{2004}, \emph{501}, 135--147.
\bibitem{hultmark_b} Hultmark, M. \emph{Private Communication}, 2017.
\bibitem{schlichting_a} Schlichting, H.; Gersten, K. \emph{Boundary-Layer Theory}, 8th ed.; Springer: Berlin, Germany, 2000.
\bibitem{zarin_a} Zarin, N.A. Measurement of non-continuum and turbulence effects on subsonic sphere drag. \emph{NASA CR-1585}, 1970.
\bibitem{fransson_a} Fransson, J.H.M; Matsubara, M.; Alfredsson, P.H. Transition induced by free-stream turbulence. \emph{J. Fluid Mech.} \textbf{2005}, \emph{527}, 1–-25.
\bibitem{hollingsworth_a} Hollingsworth, D.K.; Bourgogne, H.-A. The development of a turbulent boundary layer in high free-stream turbulence produced by a two-stream mixing layer. \emph{Exp. Thermal Fluid Science} \textbf{1995}, \emph{11}, 210--222.
\bibitem{scheichl_a} Scheichl, B.; Kluwicki, A.; Smith, F.T. Break-away separation for high turbulence intensity and large Reynolds number. \emph{J. Fluid Mech.} \textbf{2011}, \emph{670}, 260–-300.
\bibitem{ahn_a} Ahn, J.; Sparrow, E.M.; Gorman, J.M. Turbulence intensity effects on heat transfer and fluid-flow for a circular cylinder in crossflow. \emph{Int. J. Heat Mass Trans.} \textbf{2017}, \emph{113}, 613-621.
\bibitem{hansen_a} Hansen, K.S.; Barthelmie, R.J.; Jensen, L.E.; Sommer, A. The impact of turbulence intensity and atmospheric stability on power deficits due to wind turbine wakes at Horns Rev wind farm. \emph{Wind Energy} \textbf{2012}, \emph{15}, 183--196.
\bibitem{calautit_a} Calautita, J.K.; Chaudhrya, H.N.; Hughes, B.R.; Sim, L.F. A validated design methodology for a closed-loop subsonic wind tunnel. \emph{J. Wind Eng. Ind. Aerodyn.} \textbf{2014}, \emph{125}, 180--194.
\bibitem{renard_a} Renard, N.; Deck, S. A theoretical decomposition of mean skin friction generation into physical phenomena across the boundary layer. \emph{J.~Fluid Mech.} \textbf{2016}, \emph{790}, 339--367.
\bibitem{giovanetti_a} de Giovanetti, M,; Hwang, Y.; Choi, H. Skin-friction generation by attached eddies in turbulent channel flow. \emph{J.~Fluid Mech.} \textbf{2016}, \emph{808}, 511--538.
\bibitem{versteeg_a} Versteeg, H.K; Malalasekera, W. \emph{An introduction to Computational Fluid Dynamics: The Finite Volume Method}, 2nd ed.; Pearson Education Limited: Harlow, England, 2007.
\end{thebibliography}

%%%%%%%%%%%%%%%%%%%%%%%%%%%%%%%%%%%%%%%%%%
\end{document}